\documentclass[aoas,MSNbibl,nameyear,dvips]{arximspdf}
\usepackage{mathbh}
\usepackage{graphicx}

\doi{10.1214/12-AOAS570} %kopijuoti is PTS
\volume{6}
\issue{4}
\pubyear{2012}
\firstpage{1838}
\lastpage{1860}

\makeatletter

\newcommand{\mathbol}[1]{\mathbf{#1}}
\newcommand{\grbol}[1]{\bolds{#1}}

\newcommand{\rright}{\right}
\newcommand{\lleft}{\left}

\newtheorem{theorem}{Theorem}

\newproclaim{Example}{Example}

\makeatother

\begin{document}
\begin{frontmatter}

\title{Composite Gaussian process models for emulating expensive
functions\thanksref{T1}}
\runtitle{Composite Gaussian process models}

\thankstext{T1}{Supported by U.S. National
Science Foundation Grants CMMI-0448774 and DMS-10-07574.}

\begin{aug}
\author[A]{\fnms{Shan} \snm{Ba}\corref{}\ead[label=e1]{sba3@isye.gatech.edu}}
\and
\author[A]{\fnms{V. Roshan} \snm{Joseph}\ead[label=e2]{roshan@isye.gatech.edu}}
\runauthor{S. Ba and V. R. Joseph}
\affiliation{Georgia Institute of Technology}
\address[A]{School of Industrial\\
\quad and Systems Engineering\\
Georgia Institute of Technology\\
Atlanta, Georgia 30332-0205\\
USA\\
\printead{e1}\\
\hphantom{E-mail: }\printead*{e2}} %adresu isvedimo komanda gale!
\end{aug}

% HISTORY:
\received{\smonth{8} \syear{2011}}
\revised{\smonth{5} \syear{2012}}

% ABSTRACT
%
\begin{abstract}
A new type of nonstationary Gaussian process model is developed for
approximating computationally expensive functions. The new model is a
composite of two Gaussian processes, where the first one captures the
smooth global trend and the second one models local details. The new
predictor also incorporates a flexible variance model, which makes it
more capable of approximating surfaces with varying volatility.
Compared to the commonly used stationary Gaussian process model, the
new predictor is numerically more stable and can more accurately
approximate complex surfaces when the experimental design is sparse. In
addition, the new model can also improve the prediction intervals by
quantifying the change of local variability associated with the
response. Advantages of the new predictor are demonstrated using
several examples.
\end{abstract}

% KEYWORDS
%

\end{frontmatter}

%s1 #&#
\section{Introduction}\label{sec1}\label{intro}

The modern era witnesses the prosperity of computer experiments, which
play a critical role in many fields of technological development where
the traditional physical experiments are infeasible or unaffordable to
conduct. By developing sophisticated computer simulators, people are
able to evaluate, optimize and test complex engineering systems even
before building expensive prototypes. The computer simulations are
usually deterministic (no random error), yield highly nonlinear
response surfaces, and are very time-consuming to run. To facilitate
the analysis and optimization of the underlying system, surrogate
models (or emulators) are often fitted to approximate the unknown
simulated surface based on a finite number of evaluations
[\citet{Sacetal89}]. \citet{SanWilNot03} and
\citet{FanLiSud06} provide detailed reviews on the related topics.

In computer experiments, the stationary Gaussian process (GP) model is
popularly used for approximating computationally expensive simulations.
Its framework is built on modeling the computer outputs
$Y(\mathbol{x}),\mathbol{x} \in\mathbb{R}^p$, as a realization of a
stationary GP with constant mean $\mu$ and covariance function
$\sigma^2 \operatorname{cov} (Y(\mathbf{x}+\mathbf{h}),Y(\mathbol{x}))
= \sigma^2R(\mathbol{h})$, where the correlation $R(\mathbol{h})$ is a
positive semidefinite function with $R(\mathbol{0})=1$ and
$R(-\mathbol{h})=R(\mathbol{h})$. When the above assumptions are
satisfied, the corresponding predictor can be shown to be a
\textit{best linear unbiased predictor} (BLUP), in the sense that it
minimizes the mean squared prediction error. Nevertheless, many studies
in the literature have pointed out that the artificial assumption of
second-order stationarity for the GP model are more for theoretical
convenience rather than for representing reality, and they can be
easily challenged in practice. If these assumptions deviate from the
truth, the predictor is no longer optimal, and sometimes can even be
problematic [see the discussions, e.g., in \citet{Jos06},
\citet{Xioetal07}, \citet{GraLee12}].

%f1 #&#
%
\begin{figure}[b]

\includegraphics{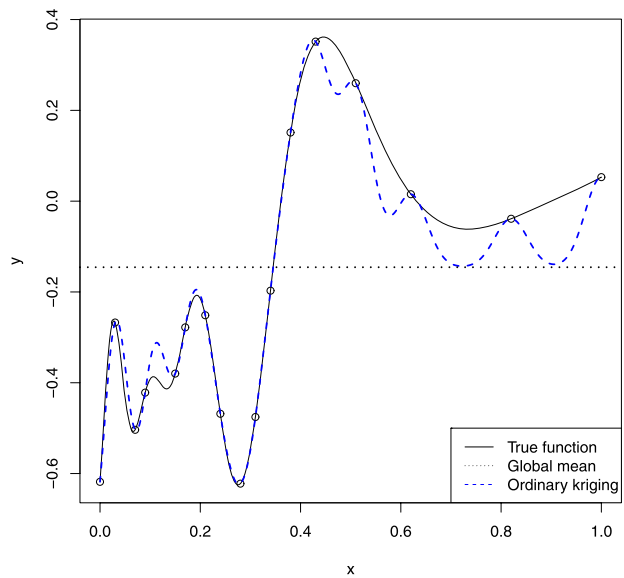}

\caption{Plot of function $y(x)=\sin(30(x-0.9)^4) \cos(2(x-0.9)) +
(x-0.9)/2$, the global mean and the ordinary kriging predictor.}
\label{figApleyOK}
\end{figure}

When the constant mean assumption for the GP model is violated, a
frequently observed consequence is that the predictor tends to revert
to the global mean, especially at locations far from design points.
Consider a simple example from \citet{Xioetal07}. Suppose the true
function is $y(x)=\sin(30(x-0.9)^4) \cos(2(x-0.9)) + (x-0.9)/2$ and
we choose 17 unequally spaced points from $[0,1]$ to evaluate the
function. The function and design points are illustrated in Figure \ref
{figApleyOK}. Obviously, the mean of this function in region $x \in
[0,0.4]$ is much smaller than the mean in region $x \in[0.4,1]$. When
the data are fitted with a stationary GP model with a Gaussian
correlation function, a constant mean for the whole region is estimated
as $-0.146$ by maximizing the likelihood function [\citet{SanWilNot03},
page 66], and the corresponding predictor along with this mean value
are shown in Figure~\ref{figApleyOK}. Clearly, the fit in region $x
\in[0.4,1]$ is not good, since the prediction is pulled down to the
global mean.

Just as a nonconstant global trend can be quite common in engineering
systems, the variability of simulated outputs can also change
dramatically throughout the design region. Still, consider the simple
case in Figure~\ref{figApleyOK}, for example: the roughness of the
one-dimensional function in region $x \in[0,0.4]$ is much larger than
in region $x \in[0.4,1]$. For the GP model assuming a constant
variance for the whole input region, the variance estimate for region
$x \in[0.4,1]$ tends to be inflated by averaging with that of the
other part, which further contributes to the erratic prediction in this
region. It is expected that as we increase the simulation sample size,
the above problem can be mitigated. However, since most typical
applications of computer experiments involve high-dimensional inputs,
the data points always tend to be sparse in the design region and it is
almost impossible to avoid such kind of gaps in practice.

In this article, we propose a more accurate modeling approach by
incorporating a flexible global trend and a variance model into the GP
model. The proposed predictor has an intuitive structure and can be
efficiently estimated in a single stage. Not only can the new predictor
mitigate the problems discussed above, it also enjoys several
additional advantages, such as better numerical stability, robustness
to sparse design and improved prediction intervals.

The article is organized as follows. Section~\ref{notation} introduces
the notation and existing work. Section~\ref{couple} presents the new
predictor and shows its interesting connections with some existing
methods. In Section~\ref{estimation} we discuss how to estimate the
unknown parameters by maximum likelihood. Several properties of the new
predictor are studied in Section~\ref{property}, and in Section \ref
{example} we use several examples to demonstrate the advantages of the
new method. Some final concluding remarks are given in Section~\ref{conclusion}.

%s2 #&#
\section{Notation and existing work}\label{sec2} \label{notation}

In the computer experiments literature, the GP model is also often
referred to as the \textit{kriging} model [\citet{Curetal91}], and
these two terms are used interchangeably in this article. Suppose we
have run the simulations under $n$ different input settings $\{
\mathbol{x}_1,\ldots,\mathbol{x}_n\} \subset\mathbb{R}^p$. Denote
the corresponding computer outputs as
$\mathbol{y}=(y_1,\ldots,y_n)^\top$. A stationary GP model, called
\textit{ordinary kriging}, can be formally stated as
%
%e1 #&#
%
\begin{equation}
\label{eqOKform} Y(\mathbol{x})=\mu+Z(\mathbol{x}),
\end{equation}
where $Z(\mathbol{x}) \sim \operatorname{GP} (0,\sigma^2 R(\cdot))$.
The ordinary kriging predictor at an input location $\mathbol{x}$ is
given by
%
%e2 #&#
%
\begin{equation}
\label{eqOKpred} \hat{y}(\mathbol{x})=\hat{\mu}+
\mathbol{r}^\top(\mathbol{x}) \mathbol{R}^{-1}(
\mathbol{y}-\hat{\mu} \mathbol{1}),
\end{equation}
where $\mathbol{r}(\mathbol{x})=(R(\mathbol{x}-\mathbol
{x}_1),\ldots,R(\mathbol{x}-\mathbol{x}_n))^\top$, $\mathbol
{R}$ is an $n \times n$ correlation matrix with the $(ij)$th element
$R(\mathbol{x}_i-\mathbol{x}_j)$, $\mathbol{1}$ is a
$n$-dimensional vector with all elements 1, and $\hat{\mu}=
(\mathbol{1}^\top\mathbol{R}^{-1}\mathbol{1})^{-1}(\mathbol
{1}^\top\mathbol{R}^{-1}\mathbol{y})$.

To remedy the predictor's reversion to mean problem as discussed in the
previous section, a common strategy is to relax the constant mean $\mu$
in ordinary kriging with a \textit{global trend} $\mu(\mathbol{x})$
and modify the model in (\ref{eqOKform}) as
%
%e3 #&#
%
\begin{equation}
\label{eqUKform} Y(\mathbol{x})=\mu(\mathbol{x})+Z(
\mathbol{x}).
\end{equation}
If the global trend is comprised of some prescribed polynomial models
$\mu(\mathbol{x})=\mathbol{f}^\top(\mathbol{x}) \grbol
{\beta}$, where $\mathbol{f}(\mathbol{x})= (1, f_1(\mathbol
{x}), \ldots, f_m(\mathbol{x}))^\top$ are known functions and
$\grbol{\beta}= (\beta_0,\beta_1, \ldots, \beta_m)^\top$ are
unknown parameters, the model in (\ref{eqUKform}) is called
\textit{universal kriging}. Define a $n\times(m+1)$ matrix $\mathbol{F} =
(\mathbol{f}(\mathbol{x}_1), \ldots, \mathbol{f}(\mathbol
{x}_n))^\top$, and the corresponding optimal predictor under model (\ref
{eqUKform}) can be derived as
%
%e4 #&#
%
\begin{equation}
\label{eqUKpred} \hat{y}(\mathbol{x})=\mathbol{f}^\top(
\mathbol{x}) \hat{\bolds{\beta}}+ \mathbol{r}^\top(
\mathbol{x}) \mathbol{R}^{-1}(\mathbol{y}- \mathbol{F}
\hat{\bolds{\beta}}),
\end{equation}
where $\hat{\bolds{\beta}} = (\mathbol{F}^\top\mathbol
{R}^{-1}\mathbol{F})^{-1}(\mathbol{F}^\top\mathbol
{R}^{-1}\mathbol{y})$. If $\mu(\mathbol{x})=
\mathbol{f}^\top(\mathbol{x}) \grbol {\beta}$ is close to the true
global trend, then clearly this approach can give much better
prediction than that of (\ref{eqOKpred}). However, in practice, the
correct functional form $\mathbol {f}(\mathbol{x})= (1,
f_1(\mathbol{x}), \ldots, f_m(\mathbol {x}))^\top$ is rarely known, and
a wrongly specified trend in universal kriging can make the prediction
even worse. For this reason, \citet{Weletal92} suggested using
ordinary kriging instead of universal kriging. Another practical
approach, called \textit{blind kriging}, is to relax the assumption
that the $f_i(\mathbol{x})$'s are known and select them from a
candidate set of functions using a variable selection technique
[\citet{JosHunSud08}]. Although this strategy usually leads to
better fit, performing the variable selection while interacting with
the second stage GP model is a nontrivial task. Considerable
computational efforts are needed to properly divide up the total
variation between the polynomial trend and the GP model. In addition,
in some cases, polynomial models may not be adequate to fit the complex
global trend well.

Generalizing the GP model for nonstationary variance is an even more
challenging task. None of the above remedies for the nonstationary mean
can in any sense alleviate the constant variance restriction, and most
studies in the literature focus on deriving complex nonstationary
covariance functions such as by spatial deformations or kernel
convolution approaches [e.g., see \citet{SamGut92},
\citet{HigSwaKer99}, \citet{SchOHa03}, \citet{PacSch06} and \citet{AndSte08}]. However, those
structures may easily get overparameterized in high dimensions and
become computationally intractable to fit. In addition, many of them
also require multiple observations, which is not applicable to the
single set of outputs from computer experiments. Some other work
includes \citet{Xioetal07}, which adopts a nonlinear mapping approach
based on a parameterized density function to incorporate the
nonstationary covariance structure. \citet{GraLee08} utilize the
Bayesian treed structure to implement a nonstationary GP model.
However, by dividing the design space into subregions, the treed GP
model may lose efficiency since the prediction is only based on local
information, and its response can also be discontinuous across
subregions. In the next section we propose to solve the nonstationarity
problem via a different approach. We show that the flexible mean and
variance models can be incorporated into GP by using the
\textit{composite Gaussian process} (CGP) models.

%s3 #&#
\section{Composite Gaussian process models}\label{sec3}\label{couple}

For clarity, in this section we develop the new method in two steps.
First, a predictor that intrinsically incorporates a flexible mean
model is presented, and then we further augment it with a variance
model to simultaneously handle the change of variability in the response.

%s3.1 #&#
\subsection{Improving the mean model}\label{sec31} \label{mean}

The universal kriging (or blind kriging) in (\ref{eqUKform}) contains
a polynomial mean model $\mu(\mathbol{x})$ as the global trend and a
kriging model $Z(\mathbol{x})$ for local adjustments. To avoid the
awkward variable selections in $\mu(\mathbol{x})$ and also make the
mean model more flexible, we propose to use another GP to model the $\mu
(\mathbol{x})$ as in the following form:
%
%e5 #&#
%
\begin{eqnarray}
\label{eqCGPform}
Y(\mathbol{x})&=&Z_{\mathrm{global}}(
\mathbol{x})+Z_{\mathrm{local}}(\mathbol{x}),
\nonumber\\
Z_{\mathrm{global}}(\mathbol{x}) &\sim& \operatorname{GP}\bigl(\mu,\tau^2g(\cdot)
\bigr),
\\
Z_{\mathrm{local}}(\mathbol{x}) &\sim& \operatorname{GP}\bigl(0,\sigma^2l(\cdot)
\bigr).
\nonumber
\end{eqnarray}
Here the two GPs $Z_{\mathrm{global}}(\mathbol{x})$ and $Z_{\mathrm{local}}(\mathbol
{x})$ are stationary and independent of each other. The first GP with
variance $\tau^2$ and correlation structure $g(\grbol{\cdot})$ is
required to be \textit{smoother} to capture the global trend, while the
second GP with variance $\sigma^2$ and correlation $l(\grbol{\cdot
})$ is for local adjustments. Just as the universal kriging generalizes
the ordinary kriging by adding a polynomial mean model $\mu(\mathbol
{x})$, the new model in (\ref{eqCGPform}) can be viewed as a further
extension which adopts a more sophisticated GP for global trend modeling.
It is interesting to note that the \textit{linear model of
regionalization} in geostatistics [\citet{Wac03}, Chapter 14] also
employs a similar structure to model regionalized phenomena in
geological data, but its final model form and estimation strategies are
quite different from our approach.

Under the new assumptions in (\ref{eqCGPform}), the optimal predictor
is easy to derive. Since the sum of two independent GPs is still a GP,
we can equivalently express (\ref{eqCGPform}) as $Y(\mathbol{x})
\sim \operatorname{GP}(\mu,\tau^2g(\cdot)+ \sigma^2l(\cdot))$. Similar to ordinary
kriging, the best linear unbiased predictor under the assumptions in
(\ref{eqCGPform}) can be written as
%
%e6 #&#
%
\begin{equation}
\label{eqCGPpred} \hat{y}(\mathbol{x})=\hat{\mu}+ \bigl(
\mathbol{g}(\mathbol{x})+ \lambda\mathbol{l}(\mathbol{x})
\bigr)^\top(\mathbol{G}+\lambda\mathbol{L})^{-1} (
\mathbol{y}- \hat{\mu}\mathbol{1}),
\end{equation}
where $\lambda=\sigma^2/\tau^2 $ $(\lambda\in[0,1])$ is the ratio of
variances,
$\mathbol{g}(\mathbol{x})=(g(\mathbol{x}-\mathbol
{x}_1),\ldots,\break g(\mathbol{x}-\mathbol{x}_n))^\top$, $\mathbol
{l}(\mathbol{x})=(l(\mathbol{x}-\mathbol{x}_1),\ldots,l(\mathbol
{x}-\mathbol{x}_n))^\top$,
$\mathbol{G}$ and $\mathbol{L}$ are two $n \times n$ correlation
matrices with the $(ij)$th element
$g(\mathbol{x}_i-\mathbol{x}_j)$ and
$l(\mathbol{x}_i-\mathbol{x}_j)$, respectively, and $\hat{\mu}=
(\mathbol{1}^\top(\mathbol{G}+\lambda\mathbol
{L})^{-1}\mathbol{1})^{-1}\mathbol{1}^\top(\mathbol{G}+\lambda
\mathbol{L})^{-1}\mathbol{y}$. Here the variance ratio $\lambda$
is restricted to $[0,1]$ because we expect the global trend to capture
most of the variation in the response surface than the local process.

Although many possible correlation structures are available for
$g(\grbol{\cdot})$ and $l(\grbol{\cdot})$, throughout this
paper we follow the standard choice in computer experiments and specify
them using the \textit{Gaussian correlation functions}:
%
%e7 #&#
%
\begin{equation}
\label{eqcorr} g(\mathbol{h})=\exp\Biggl(-\sum_{j=1}^p
\theta_j h_j^2\Biggr),\qquad l(\mathbol{h})=
\exp\Biggl(-\sum_{j=1}^p
\alpha_j h_j^2\Biggr),
\end{equation}
where $\grbol{\theta}=(\theta_1,\ldots,\theta_p)$ and $\grbol
{\alpha}=(\alpha_1,\ldots,\alpha_p)$ are unknown correlation parameters
satisfying $\mathbol{0} \le\grbol{\theta} \le\grbol
{\alpha}^l$ and $\grbol{\alpha}^l \le\grbol{\alpha}$. The
bounds $\grbol{\alpha}^l$ are usually set to be moderately large,
which ensures that the component $Z_{\mathrm{global}}(\mathbol{x})$ is indeed
smoother than $Z_{\mathrm{local}}(\mathbol{x})$ in the fitted model.

The new predictor in (\ref{eqCGPpred}) is still an interpolator,
since $\hat{y}(\mathbol{x}_i)=\hat{\mu}+ \mathbol{e}_i^\top
(\mathbol{y}- \hat{\mu}\mathbol{1})=y_i$ for $i=1,\ldots,n$,
where $\mathbol{e}_i$ is a unit vector with a 1 at its $i$th
position. It can also be seen that when $\lambda=0$ (i.e., $\sigma
^2=0$), the new model reduces to ordinary kriging. When $\lambda\in
(0,1]$, the predictor in (\ref{eqCGPpred}) can be written out as the
sum of a global predictor and a local predictor
%
%e8 #&#
%e9 #&#
%e10 #&#
%
\begin{eqnarray}
\label{eqCGPpred2} \hat{y}(\mathbol{x}) &=&
\hat{y}_{\mathrm{global}}(\mathbol{x})+\hat{y}_{\mathrm{local}}(\mathbol{x}),
\\
\label{eqCGPpred2g} \hat{y}_{\mathrm{global}}(
\mathbol{x}) &=& \hat{\mu}+ \mathbol{g}^\top(\mathbol{x}) (
\mathbol{G}+\lambda\mathbol{L})^{-1} (\mathbol{y}- \hat{\mu}
\mathbol{1}),
\\
\hat{y}_{\mathrm{local}}(\mathbol{x}) &=& \lambda\mathbol{l}^\top(
\mathbol{x}) (\mathbol{G}+\lambda\mathbol{L})^{-1} (
\mathbol{y}- \hat{\mu}\mathbol{1}).
\end{eqnarray}
It is important to note that, since the lower bounds for $\grbol
{\alpha}$ in (\ref{eqcorr}) are usually set to be moderately large,
the off-diagonal elements in $\mathbol{L}$ are closer to zero.
Particularly, we can obtain $\mathbol{L} \rightarrow\mathbol{I}$
when $\grbol{\alpha}$ take very large values. This immediately
suggests two interesting properties for the CGP model. First, its
global trend predictor $\hat{y}_{\mathrm{global}}(\mathbol{x})$ in (\ref
{eqCGPpred2g}) resembles a kriging predictor with nugget effect as
$\mathbol{L} \rightarrow\mathbol{I}$. When $\lambda>0$, this
nugget predictor is smooth but noninterpolating, and is commonly used
in spatial statistics for modeling observational data with noise
[\citet{Cre91}]. Second, since $\mathbol{L} \approx\mathbol
{I}$, the $\lambda$ in $(\mathbol{G}+\lambda\mathbol{L})$ is
mainly added to the diagonal elements. This makes $(\mathbol
{G}+\lambda\mathbol{L})$ resistent to become ill-conditioned and the
computation of $(\mathbol{G}+\lambda\mathbol{L})^{-1}$ in CGP can
be numerically very stable. These two properties are elaborated in
detail in Section~\ref{property}.

%s3.2 #&#
\subsection{Improving both the mean and variance models}\label{sec32}
\label{varsec}

To further relax the constant variance restriction, we introduce a
variance model $\sigma^{2 }(\mathbol{x})$ into (\ref{eqCGPform})
as follows:
%
%e11 #&#
%
\begin{eqnarray}
\label{eqCVGPform} Y(\mathbol{x})&=&Z_{\mathrm{global}}(
\mathbol{x})+\sigma(\mathbol{x}) Z_{\mathrm{local}}(\mathbol{x}),
\nonumber\\
Z_{\mathrm{global}}(\mathbol{x}) &\sim& \operatorname{GP}\bigl(\mu,\tau^2g(\cdot)
\bigr),
\\
Z_{\mathrm{local}}(\mathbol{x}) &\sim& \operatorname{GP}\bigl(0,l(\cdot)\bigr).
\nonumber
\end{eqnarray}
The $Z_{\mathrm{global}}(\mathbol{x})$ above remains the same as in (\ref
{eqCGPform}), since the global trend is smooth and can reasonably be
assumed to be stationary. After subtracting $Z_{\mathrm{global}}(\mathbol
{x})$ from the response, the second process is augmented with a
variance model to quantify the change of local variability such that
$\sigma(\mathbol{x}) Z_{\mathrm{local}}(\mathbol{x}) \sim \operatorname{GP}(0,\sigma
^{2}(\mathbol{x})l(\cdot))$. Overall, the model form in (\ref
{eqCVGPform}) is equivalent to assuming that the response
$Y(\mathbol{x}) \sim \operatorname{GP}(\mu,\tau^2g(\cdot)+ \sigma^{ 2}(\mathbol
{x})l(\cdot))$.

Without loss of generality, suppose the variance model can be expressed
as $\sigma^{2}(\mathbol{x})=\sigma^2 v(\mathbol{x})$, where
$\sigma^{2}$ is an unknown variance constant and $v(\mathbol{x})$ is
the standardized volatility function which fluctuates around the unit
value. In the following discussion, we first assume that $v(\mathbol
{x})$ is known and denote $\grbol{\Sigma}=\operatorname{diag}\{v(\mathbol
{x}_1),\ldots, v(\mathbol{x}_n) \}$ to represent the standardized
local variances at each of the design points $\{
\mathbol{x}_1,\ldots, \mathbol{x}_n \}$. An efficient strategy
for obtaining the $v(\mathbol{x})$ function is presented at the end
of this section.

The model assumptions in (\ref{eqCVGPform}) suggest that
$y(\mathbol{x})$ and $\mathbol{y}=(y_1,\ldots,y_n)^\top$ have the
multivariate normal distribution
%
%e12 #&#
%
\begin{eqnarray}
\label{multinormal}
 &&
\pmatrix{y(\mathbol{x})
\cr
\mathbol{y}}
\nonumber\hspace*{-20pt}\\
&&\qquad\sim N_{1+n} \lleft[
\pmatrix{\mu
\cr
\mu\mathbol{1} },\right.\hspace*{-20pt}\\
&&\qquad\hspace*{41.6pt}\left.{\fontsize{10.4pt}{11pt}\selectfont{\pmatrix{
\tau^2+\sigma^2 v(\mathbol{x}) &
\bigl(\tau^2 \mathbol{g}(\mathbol{x})+ \sigma^2
v^{1/2}(\mathbol{x})\grbol{\Sigma}^{1/2}
\mathbol{l}(\mathbol{x})\bigr)^\top
\vspace*{2pt}\cr
\tau^2 \mathbol{g}(\mathbol{x})+ \sigma^2
v^{1/2}(\mathbol{x})\grbol{\Sigma}^{1/2}
\mathbol{l}(\mathbol{x}) & \tau^2\mathbol{G}+
\sigma^2 \grbol{\Sigma}^{1/2}\mathbol{L}\grbol{
\Sigma}^{1/2} }}}\rright].
\nonumber\hspace*{-20pt}
\end{eqnarray}
The best linear unbiased predictor under these assumptions can be
derived as
%
%e13 #&#
%
\begin{eqnarray}
\label{eqCVGPpred}\quad \hat{y}(\mathbol{x})&=& \hat{\mu}+ \bigl(
\tau^2 \mathbol{g}(\mathbol{x})+ \sigma^2
v^{1/2}(\mathbol{x})\grbol{\Sigma}^{1/2}
\mathbol{l}(\mathbol{x})\bigr)^\top\bigl(\tau^2
\mathbol{G}+\sigma^2 \grbol{\Sigma}^{1/2}
\mathbol{L}\grbol{\Sigma}^{1/2}\bigr)^{-1}\nonumber\\
&&\hspace*{0pt}{}\times (
\mathbol{y}- \hat{\mu}\mathbol{1})
\\
&=& \hat{\mu}+ \bigl(\mathbol{g}(\mathbol{x})+ \lambda v^{1/2}(
\mathbol{x})\grbol{\Sigma}^{1/2} \mathbol{l}(\mathbol{x})
\bigr)^\top\bigl(\mathbol{G}+\lambda\grbol{\Sigma
}^{1/2}\mathbol{L}\grbol{\Sigma}^{1/2}
\bigr)^{-1} (\mathbol{y}- \hat{\mu}\mathbol{1}),
\nonumber
\end{eqnarray}
where $\lambda=\sigma^2/\tau^2$ $(\lambda\in[0,1])$, $\hat{\mu}=
(\mathbol{1}^\top(\mathbol{G}+\lambda\grbol{\Sigma
}^{1/2}\mathbol{L}\grbol{\Sigma}^{1/2})^{-1}\mathbol
{1})^{-1} \mathbol{1}^\top(\mathbol{G}+\break\lambda\grbol{\Sigma
}^{1/2}\*\mathbol{L}\grbol{\Sigma}^{1/2})^{-1}\mathbol{y}$ and
all the other notation remain the same as in (\ref{eqCGPpred}). Note
that after defining the ratio $\lambda$, the unknown $\sigma^2$ is no
longer needed for prediction, because the predictor depends on the
variance model $\sigma^2(\mathbol{x})$ only through $\lambda$ and
$v(\mathbol{x})$. The predictor includes (\ref{eqCGPpred})\vadjust{\goodbreak} as a
special case when the local volatility model $v(\mathbol{x})$
degenerates to a constant function. The predictor can also interpolate
all the data points since $(\mathbol{g}(\mathbol{x}_i)+ \lambda
v^{1/2}(\mathbol{x}_i)\grbol{\Sigma}^{1/2} \mathbol
{l}(\mathbol{x}_i))^\top(\mathbol{G}+\lambda\grbol{\Sigma
}^{1/2}\mathbol{L}\grbol{\Sigma}^{1/2})^{-1} =\mathbol
{e}_i^\top$ and $\hat{y}(\mathbol{x}_i)=\hat{\mu}+ \mathbol
{e}_i^\top(\mathbol{y}- \hat{\mu}\mathbol{1})=y_i$ for
$i=1,\ldots,n$.
By decomposing the predictor (\ref{eqCVGPpred}) into two parts
%
%e14 #&#
%e15 #&#
%e16 #&#
%
\begin{eqnarray}
\label{eqCVGPpred2} \hat{y}(\mathbol{x})&=&
\hat{y}_{\mathrm{global}}(\mathbol{x})+\hat{y}_{\mathrm{local}}(\mathbol{x}),
\\
\label{eqCVGPpred2g} \hat{y}_{\mathrm{global}}(
\mathbol{x})&=& \hat{\mu}+ \mathbol{g}^\top(\mathbol{x}) \bigl(
\mathbol{G}+\lambda\grbol{\Sigma}^{1/2}\mathbol{L}
\grbol{\Sigma}^{1/2}\bigr)^{-1} (\mathbol{y}- \hat{\mu}
\mathbol{1}),
\\
\hat{y}_{\mathrm{local}}(\mathbol{x}) &=& \lambda v^{1/2}(\mathbol
{x})\mathbol{l}^\top(\mathbol{x}) \grbol{
\Sigma}^{1/2} \bigl(\mathbol{G}+\lambda\grbol{
\Sigma}^{1/2}\mathbol{L}\grbol{\Sigma}^{1/2}
\bigr)^{-1} (\mathbol{y}- \hat{\mu}\mathbol{1}),
\end{eqnarray}
we can see that the global trend $\hat{y}_{\mathrm{global}}(\mathbol{x})$ in
(\ref{eqCVGPpred2g}) reduces to a \textit{stochastic kriging}
predictor [\citet{AnkNelSta10}] when $\mathbol{L}
\rightarrow\mathbol{I}$. Different from the nugget predictor in
(\ref{eqCGPpred2g}) where a universal term $\lambda$ is used for
adjusting the global trend throughout the whole region, the amount of
shrinkage at each data point in (\ref{eqCVGPpred2g}) is proportional
to the value of $\lambda v(\mathbol{x}_i)$. This \textit{localized
adjustment} scheme is advantageous in making the global trend smoother
and more stable, since it is less affected by the data points with
large variability.

The above predictor form is derived based on $Y(\mathbol{x}) \sim
\operatorname{GP}(\mu,\tau^2g(\cdot)+ \sigma^{2}(\mathbol{x})l(\cdot))$, which
unifies the modeling assumptions (\ref{eqCVGPform}) in a \textit{single
stage}. As a result, the new method can also be viewed as extending the
kriging model with a nonstationary covariance structure $\tau^2g(\cdot
)+ \sigma^{2}(\mathbol{x})l(\cdot)$. Different from this, another
strategy to fulfill the new assumptions in (\ref{eqCVGPform}) is to
develop the global and local models \textit{sequentially}: (i) Fit a
global trend model as in (\ref{eqCVGPpred2g}) using the likelihood
method. (ii) Obtain its residuals $\mathbol{s}=(\mathbol{y}-\hat
{\mathbol{y}}_{\mathrm{global}})$, where $\hat{\mathbol{y}}_{\mathrm{global}}=(\hat
{y}_{\mathrm{global}}(\mathbol{x}_1),\ldots,\hat{y}_{\mathrm{global}}(\mathbol
{x}_n))^\top$. If the estimated global trend interpolates all the data
points $(\hat{\lambda}=0)$, we have $\mathbol{s}=\mathbol{0}$ and
in this case the CGP just degenerates to a traditional single GP model.
(iii) If $\mathbol{s} \ne\mathbol{0}$, standardize the residuals
to achieve variance homogeneity $\mathbol{s}^\ast=\grbol{\Sigma
}^{-1/2} \mathbol{s}$. (iv) Adjust the global trend by interpolating
the standardized residuals via a simple kriging model $\hat
{y}_{\mathrm{adj}}(\mathbol{x})=\mathbol{l}^\top(\mathbol
{x})\mathbol{L}^{-1} \mathbol{s}^\ast$. In this way, we can form
a sequential predictor as
%
%e17 #&#
%
\begin{equation}
\label{seqpred}\qquad \hat{y}_{\mathrm{seq}}(\mathbol{x})=
\hat{y}_{\mathrm{global}}(\mathbol{x})+v^{1/2}(\mathbol{x})
\hat{y}_{\mathrm{adj}}(\mathbol{x})=\hat{y}_{\mathrm{global}}(\mathbol{x})+
v^{1/2}(\mathbol{x})\mathbol{l}^\top(\mathbol{x})
\mathbol{L}^{-1} \mathbol{s}^\ast.
\end{equation}
It is of natural interest to ask whether this sequential predictor
would make any difference from the single-stage predictor (\ref
{eqCVGPpred}), and the following theorem establishes their connections.

%th1 #&#
%
\begin{theorem}\label{th1}
Given the same parameter values, the single-stage predictor (\ref
{eqCVGPpred}) and the sequential predictor (\ref{seqpred}) are equivalent.
\end{theorem}

Proof of the theorem is left in the \hyperref[app]{Appendix}. Despite
this equivalent
model form, we want to emphasize that the single-stage fitting strategy
is superior to the sequential one in parameter estimation. This is
because all parameters in the single-stage predictor\vadjust{\goodbreak}
(\ref{eqCVGPpred}) can be optimized simultaneously, which takes into
account the interactions between global and local models and
automatically balances their effects. In contrast to this global
optimization, the sequential fitting approach estimates the parameters
in two separate steps, and each of them can at most achieve local
optimality. Generally, the global trend is hard to identify correctly
without considering the effects of the second stage model, and in many
cases the performance of the final prediction can be quite sensitive to
this ``global-local trade-off.'' As a result, in this paper we only
consider the single-stage modeling framework, and this is also a major
advantage for the proposed method over other multi-step strategies such
as blind kriging.

In the rest of this section, we present how to obtain the $v(\mathbol
{x})$ function, which is required for the CGP predictor. As shown in
(\ref{eqCVGPpred2}), the CGP model can be decomposed into a global
and a local component, and this structure provides us a convenient way
to assess the change of local volatility. For a given global trend (\ref
{eqCVGPpred2g}) (initially we can set $\grbol{\Sigma
}=\mathbol{I}$), its squared residuals $\mathbol
{s}^2=(s_1^2,\ldots,s_n^2)^\top$ are natural measures of the local
volatility, which can be used as the bases to build the $v(\mathbol
{x})$ function. Based on $\mathbol{s}^2$, we propose an intuitive
\textit{Gaussian kernel regression model} for $v(\mathbol{x})$ as
%
%e18 #&#
%
\begin{equation}
\label{varpred} v(\mathbol{x})= \frac{\mathbol{g}_b^\top(\mathbol
{x})\mathbol{s}^2}{\mathbol{g}_b^\top(\mathbol{x})\mathbol
{1}},
\end{equation}
where\vspace*{1pt} $\mathbol{g}_b(\mathbol{x})=(g_b(\mathbol{x}-\mathbol
{x}_1),\ldots,g_b(\mathbol{x}-\mathbol{x}_n))^\top$ with
$g_b(\mathbol{h})=\exp(-b\sum_{j=1}^p \theta_j h_j^2)$. Here
$\grbol{\theta}$ are the correlation parameters used in the global
trend (\ref{eqCVGPpred2g}), $b \in[0,1]$ is an extra bandwidth
parameter such that $\mathbol{g}_b(\mathbol{x}) \rightarrow
\mathbol{1}$ as $b \rightarrow0$, and $\mathbol{g}_b(\mathbol
{x})= \mathbol{g}(\mathbol{x})$ if $b=1$. Since $\mathbol
{g}(\mathbol{x})$ is the correlation of the global trend, the
underlying assumption behind (\ref{varpred}) is that whenever two
points in the global trend are strongly correlated, their variances
also tend to be more related. The bandwidth parameter $b$ adds
additional flexility in controlling the smoothness of the variance
function: when equaling zero, it smoothes out $v(\mathbol{x})$ to a
constant function even if the global trend is not flat.

From the $v(\mathbol{x})$ model in (\ref{varpred}), we can evaluate
$\hat{v}_i=v(\mathbol{x}_i)$ for $i=1,\ldots,n$ and update the
matrix $\grbol{\Sigma}=\operatorname{diag}\{\hat{v}_1,\ldots,\hat{v}_n \}$. Since
$v(\mathbol{x})$ and $\grbol{\Sigma}$ are the standardized
local volatilities, we also need to rescale them as
%
%e19 #&#
%
\begin{equation}
\grbol{\Sigma} \leftarrow\grbol{\Sigma} \Big/\Biggl(\frac{1}{n} \sum
_{i=1}^n \hat{v}_i\Biggr)
\quad\mbox{and}\quad v(\mathbol{x}) \leftarrow v(\mathbol{x})\Big/\Biggl(\frac{1}{n}
\sum_{i=1}^n \hat{v}_i
\Biggr).
\end{equation}
This standardization makes the diagonal elements of $\grbol{\Sigma
}$ have unit mean, which is essential for keeping the ratio of $\sigma
^2$ to $\tau^2$ consistent in the global trend. By plugging the updated
(and standardized) $\grbol{\Sigma}$ back into (\ref
{eqCVGPpred2g}), we can repeat the above process for a few more
times. Usually three or four iterations are sufficient to stabilize the
volatility estimates. This iterative estimation for variance is similar
in spirit to the \textit{iteratively reweighted least squares} method in
classical regression.\vadjust{\goodbreak}

Before concluding this section, we want to emphasize that the
estimation of $v(\mathbol{x})$ does not need to be separately
carried out before fitting the CGP model; instead, it can be seamlessly
nested as an inner loop in estimating the whole model. The
$v(\mathbol{x})$ function above is uniquely determined by the
unknown parameters $\grbol{\theta}$ and $b$. Since its correlation
parameter $\grbol{\theta}$ are always paired and synchronized with
that of the global trend, inclusion of this volatility function
$v(\mathbol{x})$ only adds one more parameter $b$ to the whole
model.\vspace*{-3pt}

%s4 #&#
\section{Estimation}\label{sec4} \label{estimation}

In this section we derive maximum-likelihood estimators (MLEs) for the
unknown parameters in the CGP model. As suggested at the end of
previous section, given each set of $(\lambda,\mu,\tau^2,\grbol
{\theta},\grbol{\alpha},b)$ values, $v(\mathbol{x})$ and
$\grbol{\Sigma}=\operatorname{diag}\{\hat{v}_1,\ldots,\hat{v}_n \}$ values can be
uniquely determined by nesting a small inner loop in the likelihood
function.

Based on the multivariate normal assumptions in Section~\ref{varsec},
the log-likelihood function (up to an additive constant) can be written as
\begin{eqnarray*}
&&
l\bigl(\mu,\tau^2,\sigma^2,\grbol{\theta},
\grbol{\alpha},b\bigr)
\\
&&\qquad= -\tfrac{1}{2} \log\bigl(\det\bigl(\tau^2\mathbol{G}+
\sigma^2 \grbol{\Sigma}^{1/2}\mathbol{L}\grbol{
\Sigma}^{1/2}\bigr)\bigr) \\
&&\qquad\quad{}-\tfrac
{1}{2}(\mathbol{y}- \mu
\mathbol{1})^\top\bigl(\tau^2\mathbol{G}+
\sigma^2 \grbol{\Sigma}^{1/2}\mathbol{L}\grbol{
\Sigma}^{1/2}\bigr)^{-1} (\mathbol{y}- \mu\mathbol{1}).
\end{eqnarray*}
Due to the invariant property of MLE under transformations, we can
reparameterize $\lambda=\sigma^2/\tau^2$ in the log-likelihood as
%
%e20 #&#
%
\begin{eqnarray}
\label{likelihood}
&&
l\bigl(\lambda,\mu,\tau^2,\grbol{\theta},
\grbol{\alpha},b\bigr)
\nonumber\\
&&\qquad= -\tfrac{1}{2}\bigl[n\log\bigl(\tau^2\bigr)+ \log\bigl(\det
\bigl(\mathbol{G}+ \lambda\grbol{\Sigma}^{1/2}\mathbol{L}
\grbol{\Sigma}^{1/2}\bigr)\bigr)\\
&&\qquad\quad\hspace*{16.7pt}{} +(\mathbol{y}- \mu
\mathbol{1})^\top\bigl(\mathbol{G}+ \lambda\grbol{
\Sigma}^{1/2}\mathbol{L}\grbol{\Sigma}^{1/2}
\bigr)^{-1} (\mathbol{y}- \mu\mathbol{1})/\tau^2
\bigr].
\nonumber
\end{eqnarray}
Since $\grbol{\Sigma}=\operatorname{diag}\{\hat{v}_1,\ldots,\hat{v}_n \}$ can be
known through the procedures presented in the last section, the MLEs
for $\mu$ and $\tau^2$ can be easily derived from (\ref{likelihood}) as
%
%e21 #&#
%e22 #&#
%
\begin{eqnarray}
&&\hat{\mu}(\lambda,\grbol{\theta},\grbol{\alpha},b)\nonumber\\[-8pt]\\[-8pt]
&&\qquad= \bigl(
\mathbol{1}^\top\bigl(\mathbol{G}+\lambda\grbol{\Sigma
}^{1/2}\mathbol{L}\grbol{\Sigma}^{1/2}
\bigr)^{-1}\mathbol{1}\bigr)^{-1} \bigl(
\mathbol{1}^\top\bigl(\mathbol{G}+\lambda\grbol{\Sigma
}^{1/2}\mathbol{L}\grbol{\Sigma}^{1/2}
\bigr)^{-1}\mathbol{y}\bigr),\nonumber
\\
&&
\hat{\tau}^2(\lambda,\grbol{\theta},\grbol{\alpha},b)\nonumber\\[-8pt]\\[-8pt]
&&\qquad=
\frac
{1}{n} (\mathbol{y}- \hat{\mu}\mathbol{1})^\top\bigl(
\mathbol{G}+ \lambda\grbol{\Sigma}^{1/2}\mathbol{L}
\grbol{\Sigma}^{1/2}\bigr)^{-1} (\mathbol{y}- \hat{\mu}
\mathbol{1}).\nonumber
\end{eqnarray}
After substituting these values into (\ref{likelihood}), we can obtain
the MLEs for $(\lambda,\grbol{\theta},\break\grbol{\alpha},b)$ by
minimizing the following (negative) log profile likelihood
%
%e23 #&#
%
\begin{equation}
\label{obj} \phi(\lambda,\grbol{\theta},\grbol{\alpha},b)=n\log\bigl(
\hat{\tau}^2(\lambda,\grbol{\theta},\grbol{\alpha},b)
\bigr)+\log\bigl(\det\bigl(\mathbol{G}+ \lambda\grbol{
\Sigma}^{1/2}\mathbol{L}\grbol{\Sigma}^{1/2}\bigr)
\bigr),
\end{equation}
where $\lambda\in[0,1], b\in[0,1]$, $\theta_j \in[0,\alpha^l]$ and
$\alpha_j \in[\alpha^l, \infty]$ for $j=1,\ldots,p$.\vadjust{\goodbreak}

For $p$ input variables, the above likelihood function contains $2p+2$
unknown parameters. Compared to the stationary GP model whose
likelihood contains only $p$ unknown parameters, the CGP model becomes
more difficult to estimate when the input dimension $p$ gets large. To
mitigate this disadvantage, we can further assume
%
%e24 #&#
%
\begin{equation}
\label{kappa} \alpha_j = \theta_j + \kappa,\qquad
j=1,\ldots,p.
\end{equation}
Now the CGP contains only $p+3$ unknown parameters $(\lambda,\grbol
{\theta},\kappa,b)$, whose MLEs can be obtained by minimizing
%
%e25 #&#
%
\begin{equation}
\label{obj2} \phi(\lambda,\grbol{\theta},\kappa,b)=n\log\bigl(\hat{
\tau}^2(\lambda,\grbol{\theta},\kappa,b)\bigr)+\log\bigl(\det
\bigl(\mathbol{G}+ \lambda\grbol{\Sigma}^{1/2}\mathbol{L}
\grbol{\Sigma}^{1/2}\bigr)\bigr),
\end{equation}
subject to the constraints $\lambda\in[0,1], b\in[0,1]$, $\theta_j
\in[0,\alpha^l]$ and $\kappa\in[\alpha^l, \infty]$ for $j=1,\ldots,p$.

We now\vspace*{1pt} provide a general guideline for choosing the bound $\alpha^l$.
The idea is to specify the value of $\alpha^l$ based on the
space-filling properties of the design points. Suppose the design
$D=\{\mathbol{x}_1,\ldots,\mathbol{x}_n\}$ has been standardized
into the unit region of $[0,1]^p$, and then define the following
harmonic-type average inter-point distance $d_{\mathrm{avg}}$ to measure its
space-filling properties [\citet{BaJos11}]
\[
d_{\mathrm{avg}}= \biggl( \frac{2}{n(n-1)} \sum_{1 \le i < k \le n}
\frac
{1}{d(\mathbol{x}_i,\mathbol{x}_k)^2} \biggr)^{-{1/2}},
\]
where $d(\mathbol{x}_i,\mathbol{x}_k)= \sqrt{(\sum_{j=1}^{p}
(x_{ij}-x_{kj})^2)}$. When we assume $\theta_j=\theta$ and $\alpha
_j=\alpha$ ($j=1,\ldots,p$) in the Gaussian correlation functions (\ref
{eqcorr}), correlations between points with distance $d_{\mathrm{avg}}$ are
$\exp(-\theta d_{\mathrm{avg}}^2)$ and $\exp(-\alpha d_{\mathrm{avg}}^2)$ for the global
and local processes, respectively. Because $\exp(-\alpha d_{\mathrm{avg}}^2) \le
\exp(-\alpha^l d_{\mathrm{avg}}^2) \le\exp(-\theta d_{\mathrm{avg}}^2)$, we want to
choose the bound $\alpha^l$ to restrict the correlation in the local
process to be small while ensuring that the correlation in the global
process is not too small. Although the choice is not unique, our
empirical study suggests that a good choice is to set $\exp(-\alpha^l
d_{\mathrm{avg}}^2)=0.01$, which leads to\looseness=-1
%
%e26 #&#
%
\begin{equation}\label{boundalpha}
\alpha^l = \frac{\log100}{d_{\mathrm{avg}}^2}.
\end{equation}\looseness=0
This bound is used for estimation throughout the paper.

%s5 #&#
\section{Properties}\label{sec5}\label{property}

%s5.1 #&#
\subsection{Improved prediction for sparse data sets}\label{sec51}

\begin{figure}

\includegraphics{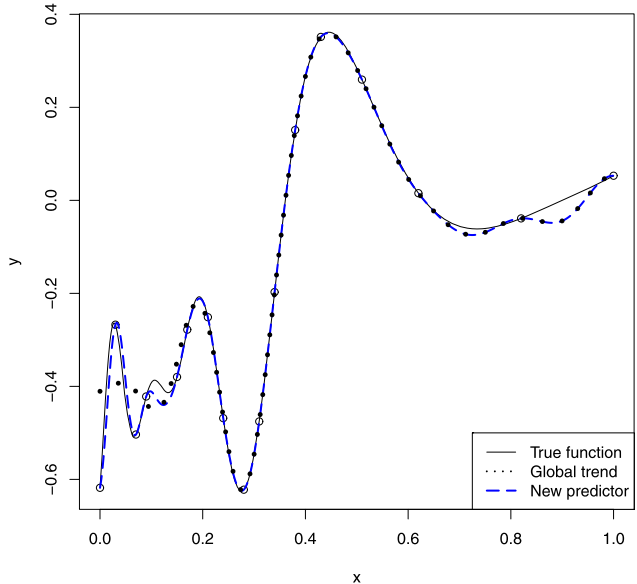}

\caption{Plot of function $y(x)=\sin(30(x-0.9)^4) \cos(2(x-0.9)) +
(x-0.9)/2$, the global trend and the CGP predictor.}
\label{figApleyCK}\vspace*{-3pt}
\end{figure}

As discussed in Section~\ref{intro}, the ordinary kriging predictor
tends to revert to the global mean in regions where data are not\vadjust{\goodbreak}
available. This erratic phenomenon will be even more pronounced if the
design points are sparse and cannot cover the input region reasonably
well. The new predictor, however, relaxes the constant mean restriction
in ordinary kriging and introduces another GP for modeling the mean.
This global trend (mean model) is noninterpolating but smooth, which
makes it immune to the erratic reversion problem in the data sparse
region. Consider again the simple test function in Figure~\ref
{figApleyOK}, where the ordinary kriging predictor ($\hat{\theta
}=469.37$) appears to be erratic. We\vspace*{2pt} fitted the CGP model ($\hat{\lambda
}=0.07, \hat{\theta}=143.6,\hat{\alpha}=1892.1, \hat{b}=1$) and its
global trend is shown as a dotted line in Figure~\ref{figApleyCK}.
Although it incurs large errors around data points in region $x \in
[0,0.4]$, it behaves well in the sparse region $[0.4,1]$ due to the
%
%f2 #&#
%
smoothness property. The final CGP predictor after incorporating the
local trend is shown as a dashed line in Figure~\ref{figApleyCK}. It
can be seen that this predictor eliminates all the noninterpolating
errors at design points. At locations far from data points, it tends to
revert to the smooth global trend instead of a global constant, which
avoids the erratic problem as in Figure~\ref{figApleyOK} and yields
much improved prediction. This shows the advantage of using the CGP
predictor when data points are sparse in some parts of the design
region. In practice, the sparseness of data points is quite common when
input dimensions are high or a nonspace-filling design is
used.\vspace*{-3pt}

%s5.2 #&#
\subsection{Numerical stability}\label{sec52}

One well-documented problem with the GP model is the potential
numerical instability when computing the inverse of its $n \times n$
correlation matrix $\mathbol{R}$.\vadjust{\goodbreak} This correlation matrix can easily
become ill-conditioned, for example, when sample size $n$ is large,
design points are close to each other, or the sample points get highly
correlated while we search for the optimal correlation parameters
[\citet{AbaBagWoo94}, \citet{HaaQia11},
\citet{PenWu}]. A near-singular correlation matrix in kriging will
lead to serious numerical problems, which causes the resulting
predictor to be unstable and unreliable.

To overcome this ill-conditioned problem, the popular approach is to
add a nonzero nugget to the diagonal elements of the correlation matrix
such that $\mathbol{R} \rightarrow(\mathbol{R}+\lambda
\mathbol{I})$. Because including a nonzero nugget has the inevitable
drawback of making predictors over-smooth (noninterpolating), in this
approach we need to reconcile the gains in numerical stability with the
losses in interpolation property and choose a trade-off value for the
nugget [\citet{RanHayKar11}, \citet{PenWu}].

As shown at the end of Section~\ref{mean}, the correlation matrix to
invert in the proposed CGP model is $(\mathbol{G}+\lambda\mathbol
{L})$. (Cases after including the variance matrix $\grbol{\Sigma}$
remain similar.) Since the lower bounds for $\grbol{\alpha}$ in
(\ref{boundalpha}) are moderately large and we have $\mathbol{L}
\approx\mathbol{I}$, the $\lambda$ in $(\mathbol{G}+\lambda
\mathbol{L})$ automatically inflates the diagonal elements of the
correlation matrix so that it is naturally resistent to becoming
singular. In addition, different from the previous nugget case, the CGP
model is always an interpolator and the $\lambda$ value here can be
freely estimated. In fact, whenever a traditional GP model has to
include a nonzero nugget for numerical reasons, the CGP model can
always improve it at least by removing its noninterpolating errors with
a augmented $Z_{\mathrm{local}}(\mathbol{x})$. This potential improvement is
shown in the next subsection.

%s5.3 #&#
\subsection{Connection with the nugget predictor}\label{sec53}

To emulate deterministic outputs from computer experiments,
\citet{GraLee12} advocate always including a nonzero nugget in the
kriging predictor for reasons even beyond computations. They argue that
when model assumptions are violated or data points are sparse, the
traditional GP predictor may lead to unpleasant results. Although
adding a nonzero nugget to the predictor incurs extra errors around
data points, it can be crucial for fitting a well-behaved (i.e.,
smooth) surface and avoiding erratic predictions in the unknown region.
In a variety of situations, \citet{GraLee12} show that overall
this noninterpolating predictor can achieve better prediction accuracy.

%f3 #&#
%
\begin{figure}

\includegraphics{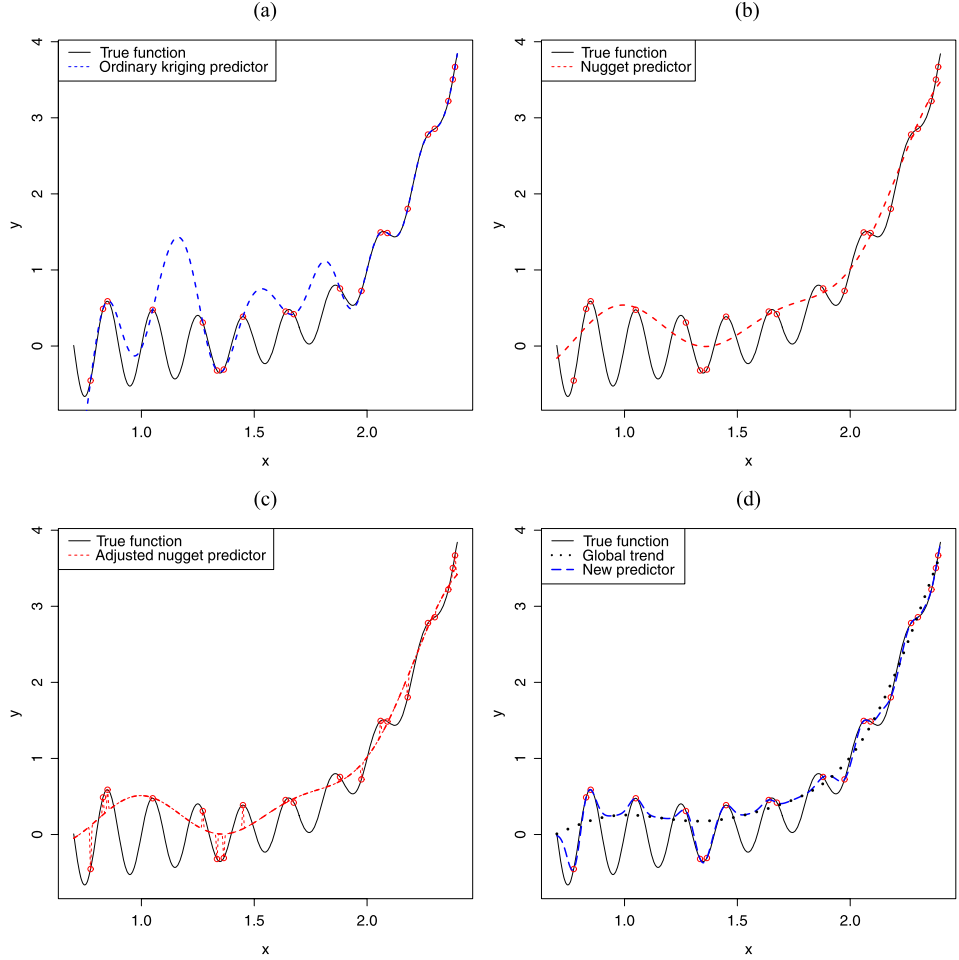}

\caption{Plot of function $y(x)=\sin(10\pi x)/(2x) +(x-1)^4$ with
\textup{(a)}
the ordinary kriging predictor; \textup{(b)} the kriging with nugget predictor;
\textup{(c)} the nugget predictor with adjustments around design points; \textup{(d)} the
optimized CGP predictor and its global trend.} \label{figGramacy}
\end{figure}

Interestingly, when the local process in CGP has zero correlation
($\mathbol{L} = \mathbol{I}$), its global trend just degenerates
to a kriging predictor with nugget, and in this case the CGP predictor
becomes $\hat{y}(\mathbol{x})= \hat{y}_{\mathrm{nugget}}(\mathbol{x})+\hat
{y}_{\mathrm{local}}(\mathbol{x})$. In regions away from design points, since
$\mathbol{l}(\mathbol{x}) = \mathbol{0}$ and $\hat
{y}_{\mathrm{local}}(\mathbol{x})=0$ for $\mathbol{x} \ne\mathbol
{x}_i$ $(i=1,2,\ldots,n)$, the CGP model exactly matches the nugget
predictor $\hat{y}_{\mathrm{nugget}}(\mathbol{x})$. At the $n$ design points,
however, due to $\mathbol{l}(\mathbol{x}) = \mathbol{e}_i$ for
$\mathbol{x} = \mathbol{x}_i$ $(i=1,2,\ldots,n)$, the $\hat
{y}_{\mathrm{local}}(\mathbol{x})$ still corrects the global trend and
adjusts the CGP to interpolate all the data points. Just as the
universal kriging generalizes the polynomial regression for\vadjust{\goodbreak}
interpolation, the CGP model can be similarly viewed as a
generalization/improvement of the nugget predictor which eliminates
errors at design points. When correlations in the local process of CGP
are further estimated as positive, the above adjustments around data
points tend to be continuous and smooth, which leads to a final CGP
predictor inheriting the advantages from both the nugget predictor and
the interpolating predictor.

Figure~\ref{figGramacy}(a) demonstrates a simulated example from
\citet{GraLee12}, where the test function $y(x)=\sin(10\pi
x)/(2x) +(x-1)^4$ is evaluated at 20 unequally spaced locations to
represent the sparseness of data points. Clearly, we can see that in
this example the ordinary kriging predictor ($\hat{\theta}=45.97$)
makes predictions well outside the range of test function in many
regions. The nugget predictor suggested by \citet{GraLee12} is
shown in Figure~\ref{figGramacy}(b). Although noninterpolating, the
nugget predictor overall gives smooth and reasonably good predictions,
which reduces the \textit{root mean squared prediction error} (RMSPE)
from the previous 0.55 to 0.35. Here the $\mbox{RMSPE}=[\frac{1}{N} \sum
_{i=1}^{N} \{\hat{y} (\mathbol{x}_i) - y(\mathbol{x}_i)\}^2
]^{1/2}$ is computed based on $N=5000$ randomly sampled data points
from the design region.
Now we further consider fitting the CGP model to this example. As shown
in Figure~\ref{figGramacy}(c), if we assume very small correlations in
$Z_{\mathrm{local}}(\mathbol{x})$, the new predictor remains almost the same
as the nugget predictor within most regions; when it comes to around
the design points, however, the predictor jumps to interpolate the
data, which slightly reduces the RMSPE to 0.34. After we also fully
estimate the correlations in $Z_{\mathrm{local}}(\mathbol{x})$ and
incorporate a variance model, Figure~\ref{figGramacy}(d) gives the
final CGP predictor ($\hat{\lambda}=0.019, \hat{\theta}=2.44,\hat{\alpha
}=578.09, \hat{b}=1$), which is smooth and gives a RMSPE as low as 0.25.

%
%s5.4 #&#
\subsection{Improved prediction intervals}\label{sec54}

\begin{figure}

\includegraphics{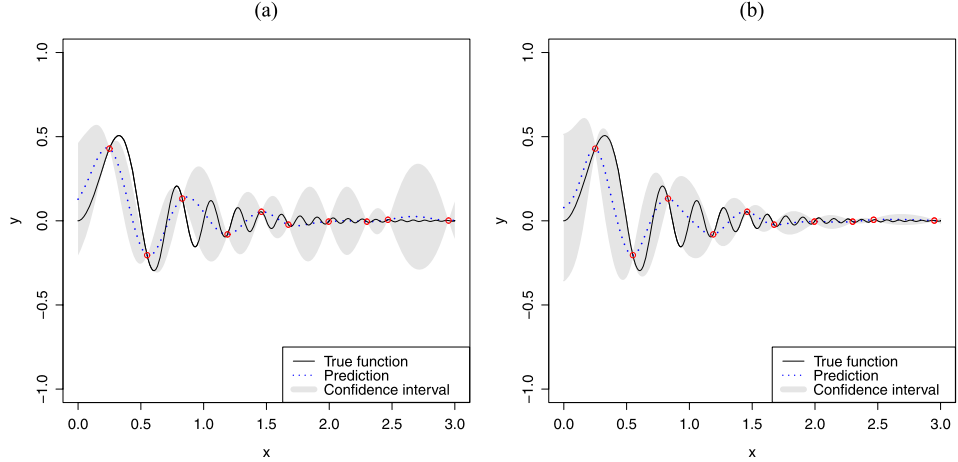}

\caption{Plot of function $y(x)=\exp(-2x)\sin(4\pi x^2)$ and the
prediction intervals from \textup{(a)} ordinary kriging;
\textup{(b)} the CGP model.}
\label{figLu}
\end{figure}

Apart from prediction, another frequently noted drawback of ordinary
kriging is the poor coverage of its prediction intervals
[\citet{Yam00}, \citet{Xioetal07}, \citet{GraLee12},
\citet{JosKan11}]. By assuming a constant variance $\sigma^2$
throughout the whole input region, the $(1-\alpha)$ prediction interval
at location $\mathbol{x}$ for ordinary kriging is given by
\[
\hat{y}(\mathbol{x}) \pm z_{\alpha/2} \sigma\biggl\{1-
\mathbol{r}^\top(\mathbol{x}) \mathbol{R}^{-1}
\mathbol{r}(\mathbol{x})+\frac
{(1 - \mathbol{r}^\top(\mathbol{x}) \mathbol{R}^{-1}
\mathbol{1} )^2}{\mathbol{1}^\top\mathbol{R}^{-1}\mathbol
{1}} \biggr\}^{1/2},
\]
where $z_{\alpha/2}$ is the upper $\alpha/2$ critical value of the
standard normal distribution. This prediction interval is often too
restrictive and inadequate to cover some complex underlying surfaces
since it fails to take into account the change of local variability in
the design region. One typical example is demonstrated in Figure
\ref{figLu}(a), where the test function fluctuates around zero with
decreasing amplitude. The corresponding prediction intervals from
ordinary kriging, however, yield the same variability pattern
throughout the whole design region, which are obviously too narrow to
cover the high volatility region in the left part, but also end up
unnecessarily wide in the right part of the input region where the true
function is almost flat. In this subsection, we introduce the
prediction intervals for CGP models. By relaxing the constant variance
restriction, these prediction intervals are self-adjusted according to
the local variability and can be expected to give much improved coverage.

%f4 #&#
%

In a Bayesian framework, the assumptions for a CGP model in (\ref
{eqCVGPform}) can be viewed as putting a prior distribution
$y(\mathbol{x})|\mu\sim \operatorname{GP}(\mu,\tau^2g(\cdot)+ \sigma^{2}(\mathbol
{x})l(\cdot))$ on the function, which leads to the
first-stage conditional distribution
\[
\pmatrix{ y(\mathbol{x})
\cr
\mathbol{y} }
\bigg\vert\mu\sim N_{1+n}
\biggl[ \pmatrix{\mu
\cr
\mu\mathbol{1} },
\tau^2
\pmatrix{
1+\lambda v(\mathbol{x}) &
\mathbol{q}^\top(\mathbol{x})
\cr
\mathbol{q}(\mathbol{x}) & \mathbol{Q} }
\biggr],
\]
where $\lambda=\sigma^2/\tau^2$, $ \mathbol{q}(\mathbol
{x})=\mathbol{g}(\mathbol{x})+ \lambda v^{1/2}(\mathbol
{x})\grbol{\Sigma}^{1/2} \mathbol{l}(\mathbol{x})$,
$\mathbol{Q}=\mathbol{G}+\lambda\grbol{\Sigma
}^{1/2}\mathbol{L}\grbol{\Sigma}^{1/2}$ and all the other
notation remains the same as in Section~\ref{varsec}. Here, for
simplicity, the variance and correlation parameters are assumed to be
known. If we further assume a second-stage noninformative prior for
$\mu\dvtx p( \mu) \propto1$ and integrate it out, then the predictive
distribution for $y(
\mathbol{x})$ can be derived as %
\[
y(\mathbol{x})|
\mathbol{y} \sim N_{1} \bigl(\mu_{0|n}(\mathbol
{x}),v^2_{0|n}(\mathbol{x})\bigr),
\]
where
\[
\mu_{0|n}(\mathbol{x}) = \hat{\mu}+ \mathbol{q}^\top(
\mathbol{x}) \mathbol{Q}^{-1} (\mathbol{y}- \hat{\mu}
\mathbol{1}) \qquad\mbox{for } \hat{\mu}= \bigl(\mathbol{1}^\top
\mathbol{Q}^{-1}\mathbol{1}\bigr)^{-1} \bigl(
\mathbol{1}^\top\mathbol{Q}^{-1}\mathbol{y}\bigr)
\]
and
%
%e27 #&#
%
\begin{equation}
\label{eqPIV} v^2_{0|n}(\mathbol{x}) =
\tau^2 \biggl\{ 1+ \lambda v(\mathbol{x})-\mathbol{q}^\top(
\mathbol{x}) \mathbol{Q}^{-1}\mathbol{q}(\mathbol{x})+
\frac{(1 - \mathbol{q}^\top(\mathbol{x})
\mathbol{Q}^{-1} \mathbol{1} )^2}{\mathbol{1}^\top\mathbol
{Q}^{-1}\mathbol{1}}\biggr\}.
\end{equation}
The derivation for these results is tedious but standard, which follows
similar development steps as in \citet{SanWilNot03}, Section 4.3.
It can be seen that our previously proposed predictor in (\ref
{eqCVGPpred}) is nothing but the posterior mean of the function given
the data. Now
a (pointwise) prediction interval for this predictor can be constructed by
%
%e28 #&#
%
\begin{equation}
\label{eqPICK} \hat{y}(\mathbol{x}) \pm z_{\alpha/2} v_{0|n}(
\mathbol{x}),
\end{equation}
where $z_{\alpha/2}$ is the upper $\alpha/2$ critical value of the
standard normal distribution.

Note that, since
$\mathbol{q}^\top(\mathbol{x}_i)\mathbol{Q}^{-1}=\mathbol
{e}_i^\top$
and $\mathbol{e}_i^\top\mathbol{q}(\mathbol{x}_i)=1+ \lambda
v(\mathbol{x}_i)$, the above posterior variance
$v^2_{0|n}(\mathbol{x})$ equals zero whenever
$\mathbol{x}=\mathbol{x}_i$ for $i=1,\ldots,n$. Thus, as in
ordinary kriging, the width of the prediction interval shrinks to zero
at each data point, which is quite intuitive since both models
interpolate the responses at each observed location. On the other hand,
however, different from ordinary kriging, the variance of predictive
distribution in (\ref{eqPIV}) depends on the local variability of the
underlying surface, which intrinsically adjusts the widths of the
prediction interval. Consider again the test function in
Figure~\ref{figLu}. It can be seen in Figure~\ref{figLu}(b) that the
prediction intervals from a CGP model ($\hat{\theta}=2.1,
\hat{\alpha}=54.85, \hat{\lambda}=1, \hat{b}=1$) become much wider in
the left region when the function fluctuates rapidly, but quickly
narrow down as the underlying function becomes flat. Compared with the
prediction intervals for ordinary kriging, the new intervals can more
precisely demonstrate the change of prediction uncertainties throughout
the input region, that is, the predictive variances are much larger in
the left part of region than in the right. One way to quantify such
improvements is through computing the \textit{interval score} for central
prediction intervals [\citet{GneRaf07}] which is defined as
$S_\alpha^{\mathrm{int}}(l,u;x)=(u-l)+\frac{2}{\alpha}(l-x)\mathbh{1}\{x<l\}
+\frac{2}{\alpha}(x-u)\mathbh{1}\{x>u\}$
for a $(1-\alpha)\%$ central prediction interval $[l,u]$. This scoring
rule (to be minimized) rewards narrow prediction intervals and also
penalizes lack of coverage. For the prediction intervals in
Figure~\ref{figLu}, the average interval score (based on 3000 randomly
sampled test points) for the ordinary kriging in (a) is 0.62 while for
the CGP model in (b) is only 0.32, which shows almost 50\% improvement.

%s5.5 #&#
\subsection{Extensions to noisy data}\label{sec55}

In the previous sections we model the deterministic outputs from a
computer experiment by coupling two GPs. As an extension to this,
sometimes it is also possible to use the sum of more than two GPs for
gaining additional flexibility in the model and satisfying special
needs. One important application of this extension is to modify the new
predictor for modeling data with random errors.

Based on the previous model form in Section~\ref{varsec}, we can add a
third GP (with zero correlation) to account for the white noise as follows:
\[
Y(\mathbol{x})=Z_{\mathrm{global}}(\mathbol{x})+\sigma(\mathbol{x})
Z_{\mathrm{local}}(\mathbol{x}) + \varepsilon(\mathbol{x}),
\]
where $Z_{\mathrm{global}}(\mathbol{x})$, $Z_{\mathrm{local}}(\mathbol{x})$ are the
same stationary GPs as in (\ref{eqCVGPform}), and the error term
$\varepsilon(\mathbol{x})$ is assumed to be $N(0,\sigma^2_{\varepsilon
}(\mathbol{x}))$ distributed, uncorrelated at
different input locations and also independent of the other two GPs.
Suppose the error variances $\grbol{\Sigma}_\varepsilon= \operatorname{diag}\{
\sigma^2_{\varepsilon}(\mathbol{x}_1),\ldots,\sigma^2_{\varepsilon
}(\mathbol{x}_n)\}$ are given, then the best linear unbiased
predictor can be easily updated by modifying (\ref{eqCVGPpred}) as follows:
\begin{eqnarray*}
\hat{y}(\mathbol{x}) &=& \hat{\mu}+ \bigl(\tau^2 \mathbol{g}(
\mathbol{x})+ \sigma^2 v^{1/2}(\mathbol{x})
\grbol{\Sigma}^{1/2} \mathbol{l}(\mathbol{x})
\bigr)^\top\bigl(\tau^2 \mathbol{G}+\sigma^2
\grbol{\Sigma}^{1/2}\mathbol{L}\grbol{\Sigma}^{1/2}
+\grbol{\Sigma}_\varepsilon\bigr)^{-1} \\
&&{}\times(\mathbol{y}- \hat{\mu
}\mathbol{1})
\\
&=& \hat{\mu}+ \bigl(\mathbol{g}(\mathbol{x})+ \lambda v^{1/2}(
\mathbol{x})\grbol{\Sigma}^{1/2} \mathbol{l}(\mathbol{x})
\bigr)^\top\bigl(\mathbol{G}+\lambda\grbol{\Sigma
}^{1/2}\mathbol{L}\grbol{\Sigma}^{1/2} + \rho
\grbol{\Sigma}_\varepsilon\bigr)^{-1} (\mathbol{y}- \hat{\mu}
\mathbol{1}),
\end{eqnarray*}
where $\rho=1/\tau^2$, $\hat{\mu}= (\mathbol{1}^\top(\mathbol
{G}+\lambda\grbol{\Sigma}^{1/2}\mathbol{L}\grbol{\Sigma
}^{1/2} + \rho\grbol{\Sigma}_\varepsilon)^{-1}\mathbol
{1})^{-1} (\mathbol{1}^\top(\mathbol{G}+\lambda\grbol{\Sigma
}^{1/2}\*\mathbol{L}\grbol{\Sigma}^{1/2} + \rho\grbol{\Sigma
}_\varepsilon)^{-1}\mathbol{y})$ and all the other notation remains
the same as in (\ref{eqCVGPpred}). This predictor for noisy data is
no longer an interpolator, and its parameter estimation can be
similarly carried out as in the previous sections, except for
$(\mathbol{G}+\lambda\grbol{\Sigma}^{1/2}\mathbol
{L}\grbol{\Sigma}^{1/2})$ replaced by $(\mathbol{G}+\lambda
\grbol{\Sigma}^{1/2}\mathbol{L}\grbol{\Sigma}^{1/2} + \rho
\grbol{\Sigma}_\varepsilon)$ in the models.

%s6 #&#
\section{Examples}\label{sec6} \label{example}

\begin{Example}\label{Example1}
For any nonstationary modeling approach, one
commonly raised concern is that if the true surface is indeed a
realization from a stationary Gaussian process, whether the
``unnecessarily sophisticated'' nonstationary modeling approach can
perform as good as the ``correct'' stationary model. To test the
performance of our proposed model in such cases, we simulate sample
paths from various two-dimensional stationary Gaussian processes 50
times and fit both the CGP and the stationary GP models to each of them
for comparison. A 24-run maximin distance Latin Hypercube Design (LHD)
is used in these simulations and for each time the true correlation
parameters in GP are randomly generated from $[1,5]$. In each
iteration, once the design and correlation parameters are fixed, a $24
\times24$ correlation matrix $\mathbol{R}$ is uniquely determined.
A sample path from the corresponding stationary GP can then be drawn by
simulating a random sample vector from the multivariate normal
distribution $N_n(\mu\mathbol{1}^n, \sigma^2 \mathbol{R}^n)$
with $n=24, \mu=0, \sigma^2=1$.

After drawing stationary sample paths as above 50 times, we fit CGP
models to each of them. Among the 50 fitted models, 42 out of them have
$\hat{\lambda}=0$, which shows that the CGP has perfectly degenerated
to the stationary GP model. For the other eight CGP models, their $\hat
{\lambda}$ values are also extremely small, with the largest one only
as 0.003. Measured by the leave-one-out cross-validation error, the
prediction accuracy of the CGP model and the stationary GP model are
almost identical in these cases.
\end{Example}
\begin{Example}\label{Example2}
In this example, we provide two test functions
possessing nonstationary features: one in two dimensions and the other
in 10 dimensions. The first function is $f(x_1,x_2)=\sin(1/(x_1x_2))$
$(x_1,x_2 \in[0.3,1])$, whose surface fluctuates rapidly when $x_1$ or
$x_2$ is small, but gradually becomes smooth as $x_1$ and $x_2$
increase toward one. The second test function (known as Michalewicz's
function) has the following form:
\[
f(\mathbol{x})=-\sum_{i=1}^{10}
\sin(x_i)\biggl[\sin\biggl(\frac{ix_i^2}{\pi
}\biggr)
\biggr]^{2m},\qquad 0 \le x_i \le\pi,\qquad i=1, \ldots, 10.
\]
Typically, this function is used with $m=10$, which leads to a
high-dimen\-sional surface containing many local optima, and its
volatility varies dramatically throughout the input region.

%t1 #&#
%
\begin{table}
\tablewidth=280pt
\caption{RMSPE values for the two-dimensional function in Example
\protect\ref{Example2}}\label{xiongdata2d}
\begin{tabular*}{\tablewidth}{@{\extracolsep{\fill}}l c c@{}}
\hline
\textbf{Method} & \textbf{Maximin LHD} & \textbf{Adaptive design} \\
\hline
GP & 0.188 & 0.266 \\
CGP & 0.144 & 0.159 \\
TGP & 0.312 & 0.465 \\
\hline
\end{tabular*}
\end{table}

We use a 24-run maximin distance LHD and a 24-run adaptive design from
\citet{Xioetal07} to evaluate the first test function. Both the GP\vadjust{\goodbreak}
and CGP models are fitted to these two designs, and their RMSPEs are
compared based on additional 5000 randomly sampled testing data. From
the results in Table~\ref{xiongdata2d}, we can see that the CGP
predictor improves the accuracy of the GP predictor by $23\%$ and $40\%
$ for each design. Table~\ref{xiongdata2d} also shows the results of
fitting the Bayesian treed Gaussian process (TGP) model
[\citet{GraLee08}]. The RMSPEs of this nonstationary treed model
are relatively large, which probably are due to its inefficient
partitioning of the input region.

%f5 #&#
%
\begin{figure}[b]

\includegraphics{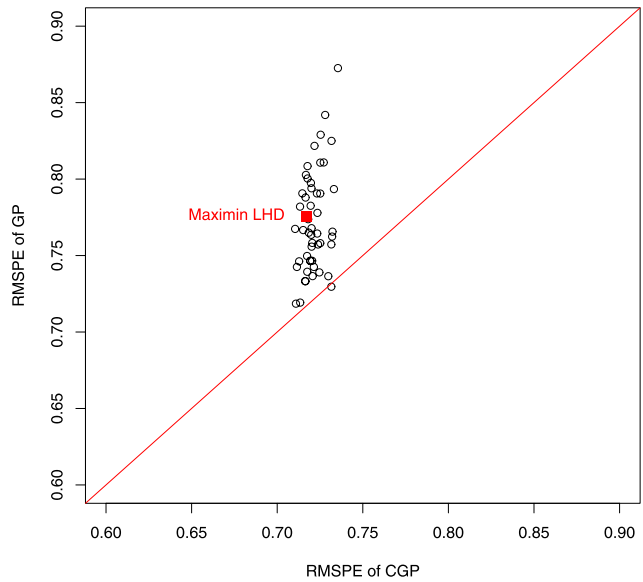}

\caption{RMSPEs of GP and CGP models for Michalewicz's function in
Example \protect\ref{Example2}. Points falling above the diagonal line
indicate larger
prediction errors for the GP model.}
\label{figMich}
\end{figure}

To further test the performance of the CGP predictor based on different
designs, we generate fifty 100-run random LHDs to evaluate the second
test function and fit the GP and CGP models to each of them. RMSPEs of
the two predictors are plotted in Figure~\ref{figMich} for the 50
random designs. It can be seen that, compared to the GP model, the CGP
predictor can always give better approximations to this complex
surface. The RMSPEs of the two predictors based on a 100-run maximin
distance LHD are also marked in this plot.
\end{Example}
\begin{Example}\label{Example3}
\citet{Qiaetal06} described a computer simulation
of a heat exchanger for electronic cooling applications. The device
under study consists of linear cellular materials and is used for
dissipating the heat generated by some sources such as a
microprocessor. The response of interest is the total rate of steady
state heat transfer of the device, which depends on the mass flow rate
of entry air $\dot{m} \in(0.00055, 0.001)$, the temperature of entry
air $T_{\mathrm{in}} \in(270, 303.15)$, the solid material thermal conductivity
$k \in(330, 400)$ and the temperature of the heat source $T_{\mathrm{wall}} \in
(202.4, 360)$. The device is assumed to have fixed overall width~(W),
depth (D) and height (H) of 9, 25 and 17.4 millimeters, respectively.
In \citet{Qiaetal06}, the study involved two types of simulators: an
expensive finite element simulator and a relatively cheaper finite
difference simulator. Since the latter type of simulation was
systematically conducted in the design space while the previous one was
only available at limited locations, here we only focus on using the
finite difference simulation results to compare the prediction accuracy
of several different models. Because the four input variables are in
very different scales, all of them are standardized into the $(0,1)$
region before analysis.

%(2006).}

\citet{Qiaetal06} used a 64-run orthogonal array-based Latin Hypercube
design for running the finite difference simulations with an extra
14-run test data set for assessing the predictions from the surrogate
model. If no prior information is available for the function and an
ordinary kriging with Gaussian correlation function is directly fitted,
the maximum likelihood estimates for its correlation parameters are
(0.22, 4.37, 0.14, 7.24), which yield a RMSPE of 5.15. However, for
this particular problem, the physical domain knowledge indicates that a
linear component is very likely to exist between the response and
factors. As a result, \citet{Qiaetal06} included the linear trend into
the model and fitted a universal kriging to the data. Their results
showed that the linear effects for $T_{\mathrm{in}}$ and $T_{\mathrm{wall}}$ are
significant but for the other two variables are almost negligible. By
including these two linear effects into the global trend, the RMSPE can
be successfully reduced to only 2.588. Now we fit a CGP model to the
data for comparison. Based on the maximum likelihood method in
Section~\ref{estimation}, we can estimate the unknown parameters as
$\hat{\grbol{\theta}}=(0.008,0.3,0.01,11.74)$, $\hat{\grbol
{\alpha}}=(11.81,12.17,11.94,23.48)$, $\hat{\lambda}=0.019$ and $\hat
{b}=1$. The RMSPE for this new predictor is 2.24, which is much better
than the ordinary kriging and even smaller than the previous improved
result from universal kriging. Note that in the global trend of this
new predictor, the two correlation parameters $\hat{\theta}_2$ and $\hat
{\theta}_4$ (for $T_{\mathrm{in}}$ and $T_{\mathrm{wall}}$) are remarkably larger than
the others, which perfectly coincides with the two significant linear
trends in universal kriging. This demonstrates the effectiveness of the
CGP model for capturing the global trend. In most common situations
where no functional relationship\vadjust{\goodbreak} in the global trend can be known in
advance, the ability to automatically estimate the trend and the
variance is a great advantage for the new predictor over the other methods.
\end{Example}

%s7 #&#
\section{Conclusions}\label{sec7}\label{conclusion}

In this article we present an intuitive approach for approximating
complex surfaces that are not second-order stationary. The new
predictor intrinsically incorporates a global trend and a flexible
variance model, and all of its parameters can be estimated in a single
stage. Compared with many existing methods, the new model enjoys
several advantages such as numerical stability, improved prediction
accuracy and flexible prediction intervals. %R codes for fitting the CGP
%model can be obtained from the authors' website.
An R package \texttt{CGP} for fitting the CGP model can be downloaded
from \url{http://www.cran.r-project.org/}.

For modeling the nonstationarity in variance, one reviewer draws our
attention to a related idea called \textit{scaling} in the geostatistical
literature [\citet{BanChaGel03}]. The scaling
approach is given in the form $Y(\mathbol{x})=\sigma(\mathbol
{x})Z(\mathbol{x})$, where $Z(\mathbol{x})$ denotes a stationary
process and $\sigma^2(\mathbol{x})$ is a variance function that
needs to be specified. By choosing $\sigma^2(\mathbol{x})$ as the
exponent of another Gaussian process, \citet{Huaetal11}
proposed a \textit{stochastic heteroscedastic process} (SHP) model
$y(\mathbol{x})=\mathbol{g}^\top(\mathbol{x}) \grbol{\beta
} +\sigma\exp(\tau\alpha(\mathbol{x})/2) Z(\mathbol{x})$ for
low-dimensional environmental applications, where $\alpha(\mathbol
{x})$ is defined to be another stationary Gaussian process that is
independent of $Z(\mathbol{x})$. Although this SHP model does not
have a flexible global trend, its variance model is more sophisticated
than our CGP model. This additional flexibility in variance, however,
comes with the expenses of a very difficult and complicated estimation
procedure. Since the likelihood function of the SHP model has no
closed-form expression, simulation-based approximations have to be
applied for the likelihood value during each step of its optimization.
Obviously, this can be computationally very challenging (or even
infeasible) when the dimension of unknown parameters is high, which
limits its application in computer experiments.

Recently, we also noticed an interesting work from
\citet{HaaQia11}, which uses the sum of multiple GPs to emulate
outputs from large scale computer experiments. However, the purposes of
their work is different from ours. The aim of \citet{HaaQia11} is
mainly to control the numerical error in computing interpolators based
on a huge amount of data. Their multiple GP models are fitted
sequentially and each of them is only based on a subset of data points.
On the contrary, our method is developed to improve the precision in
modeling expensive simulation results that are not second-order
stationary. Both our global and local GPs are fitted based on the
entire data set and all parameters in our model are also estimated in a
single stage.

For $p$ input factors, the proposed CGP model involves $p+3$ unknown
parameters, which is computationally slightly more expensive to fit
than the ordinary kriging. This is the price we need to pay for
incorporating the extra flexility in modeling the global trend and the
change of variance. We want to note that although the number of
parameters in ordinary kriging can also be extended from $p$ to $2p$ by
generalizing its Gaussian correlation function to the \textit{power
exponential correlation function} $r(\mathbol{h})=\exp(-\sum_{j=1}^p
\theta_j |h_j|^{w_j})$ or even a \textit{Matern correlation function},
this extension alone cannot solve the problems discussed in this paper,
since the resulting predictor still remains second-order stationary.

%apA #&#
%
\begin{appendix}\label{app}
\section*{\texorpdfstring{Appendix: Proof of Theorem \lowercase{\protect\ref{th1}}}{Appendix: Proof of Theorem 1}}

Since both the single-stage predictor (\ref{eqCVGPpred2}) and the
sequential predictor (\ref{seqpred}) contain the same global trend
$\hat{y}_{\mathrm{global}}(\mathbol{x})$ as in (\ref{eqCVGPpred2g}), we
only need to prove $\hat{y}_{\mathrm{local}}(\mathbol{x})=v^{1/2}(\mathbol
{x})\hat{y}_{\mathrm{adj}}(\mathbol{x})$:
\begin{eqnarray*}
&&
v^{1/2}(\mathbol{x})\hat{y}_{\mathrm{adj}}(\mathbol{x}) \\
&&\qquad=
v^{1/2}(\mathbol{x})\mathbol{l}^\top(\mathbol{x})
\mathbol{L}^{-1} \mathbol{s}^\ast
\\
&&\qquad= v^{1/2}(\mathbol{x})\mathbol{l}^\top(\mathbol
{x})\mathbol{L}^{-1} \grbol{\Sigma}^{-1/2} \bigl[
\mathbol{y}- \hat{\mu}\mathbol{1}- \mathbol{G} \bigl(\mathbol{G}+
\lambda\grbol{\Sigma}^{1/2}\mathbol{L}\grbol{
\Sigma}^{1/2}\bigr)^{-1} (\mathbol{y}- \hat{\mu}
\mathbol{1})\bigr]
\\
&&\qquad= v^{1/2}(\mathbol{x})\mathbol{l}^\top(\mathbol
{x})\mathbol{L}^{-1} \grbol{\Sigma}^{-1/2} \bigl[
\mathbol{I}-\mathbol{G}\bigl(\mathbol{G}+\lambda\grbol{\Sigma
}^{1/2}\mathbol{L}\grbol{\Sigma}^{1/2}
\bigr)^{-1}\bigr] (\mathbol{y}- \hat{\mu}\mathbol{1})
\\
&&\qquad= \lambda v^{1/2}(\mathbol{x})\mathbol{l}^\top(
\mathbol{x}) \grbol{\Sigma}^{1/2} \bigl(\lambda\grbol{
\Sigma}^{1/2}\mathbol{L}\grbol{\Sigma}^{1/2}
\bigr)^{-1} \\
&&\qquad\quad{}\times\bigl[\mathbol{I}-\mathbol{G}\bigl(\mathbol{G}+
\lambda\grbol{\Sigma}^{1/2}\mathbol{L}\grbol{
\Sigma}^{1/2}\bigr)^{-1}\bigr] (\mathbol{y}- \hat{\mu}
\mathbol{1})
\\
&&\qquad=^{(\ast)} \lambda v^{1/2}(\mathbol{x})\mathbol{l}^\top
(\mathbol{x}) \grbol{\Sigma}^{1/2} \bigl(\mathbol{G}+\lambda
\grbol{\Sigma}^{1/2}\mathbol{L}\grbol{\Sigma}^{1/2}
\bigr)^{-1} (\mathbol{y}- \hat{\mu}\mathbol{1})
\\
&&\qquad= \hat{y}_{\mathrm{local}}(\mathbol{x}),
\end{eqnarray*}
where the $\mbox{equality} =^{(\ast)}$ holds because $(\lambda\grbol
{\Sigma}^{1/2}\mathbol{L}\grbol{\Sigma}^{1/2})^{-1} [\mathbol
{I}-\mathbol{G}(\mathbol{G}+\lambda\grbol{\Sigma
}^{1/2}\*\mathbol{L}\grbol{\Sigma}^{1/2})^{-1}] (\mathbol
{G}+\lambda\grbol{\Sigma}^{1/2}\mathbol{L}\grbol{\Sigma
}^{1/2}) = \mathbol{I}$.
\end{appendix}

\section*{Acknowledgments}

The authors thank the Editor and two referees for their valuable
comments and suggestions.

%suskaldyti doi

% imsref loaded by lrinkeviciute, 2012-07-05 13:34:22
% imsref loaded by lrinkeviciute, 2012-07-05 13:40:42

\printaddresses

\end{document}